\documentclass[aps,twocolumn, superscriptaddress, svgnames]{revtex4-2}
\usepackage[T1]{fontenc} % Use modern font encodings
\usepackage{tikz,tikzscale, relsize, pgfplots, xcolor, amssymb, amsmath, gensymb}

\usetikzlibrary{backgrounds, arrows, calc, decorations, positioning, decorations.pathmorphing,decorations.pathreplacing, 
	decorations.markings, fixedpointarithmetic, fit, patterns, decorations.text, backgrounds,matrix,plotmarks, spy}
\usepgfplotslibrary{fillbetween}
\pgfplotsset{compat=1.17}
\usetikzlibrary{external}
\usepackage{graphicx}
\graphicspath{{images/}}
\usepackage[percent]{overpic}

\newcommand{\vect}[1]{{\boldsymbol{#1}}}

\begin{document}
\title{Microscopic Observation of Non-Ergodic States in Two-Dimensional Non-Topological Bubble Lattices}

\author{S. Pylypenko*}
\email{s.pylypenko@ifw-dresden.de}
\affiliation{Leibniz Institute for Solid State and Materials Research Dresden, Helmholtzstraße 20, 01069 Dresden, Germany}
\author{M. Winter*}
\email{Moritz.Winter@cpfs.mpg.de}
\affiliation{Max Planck Institute for Chemical Physics of Solids, 01187 Dresden, Germany}
\affiliation{Dresden Center for Nanoanalysis, cfaed, TUD Dresden University of Technology, 01069 Dresden, Germany.}
\author{U. K. R\"{o}\ss{}ler}
\affiliation{Leibniz Institute for Solid State and Materials Research Dresden, Helmholtzstraße 20, 01069 Dresden, Germany}
\author{D. Pohl}
\affiliation{Dresden Center for Nanoanalysis, cfaed, TUD Dresden University of Technology, 01069 Dresden, Germany.}
\author{R. Kyrychenko}
\affiliation{Leibniz Institute for Solid State and Materials Research Dresden, Helmholtzstraße 20, 01069 Dresden, Germany}
\author{M. C. Rahn}
\affiliation{Institute for Solid State and Materials Physics, Technical University of Dresden, 01062 Dresden, Germany.}
\affiliation{Experimental Physics VI, Center for Electronic Correlations and Magnetism, Institute of Physics, University of Augsburg, 86159 Augsburg, Germany}
\author{B. Achinuq}
\affiliation{Department of Physics, Clarendon Laboratory, University of Oxford, Oxford, OX1~3PU, United Kingdom}
\author{J. R. Bollard}
\affiliation{Department of Physics, Clarendon Laboratory, University of Oxford, Oxford, OX1~3PU, United Kingdom}
\affiliation{Diamond Light Source, Harwell Science and Innovation Campus, Didcot, OX11~0DE, United Kingdom}
\author{P. Vir}
\affiliation{Max Planck Institute for Chemical Physics of Solids, 01187 Dresden, Germany}
\author{G. van der Laan}
\affiliation{Diamond Light Source, Harwell Science and Innovation Campus, Didcot, OX11~0DE, United Kingdom}
\author{T. Hesjedal}
\affiliation{Department of Physics, Clarendon Laboratory, University of Oxford, Oxford, OX1~3PU, United Kingdom}
\affiliation{Diamond Light Source, Harwell Science and Innovation Campus, Didcot, OX11~0DE, United Kingdom}
\author{J. Schultz}
\affiliation{Leibniz Institute for Solid State and Materials Research Dresden, Helmholtzstraße 20, 01069 Dresden, Germany}
\author{B. Rellinghaus}
\affiliation{Dresden Center for Nanoanalysis, cfaed, TUD Dresden University of Technology, 01069 Dresden, Germany.}
\author{C. Felser}
\affiliation{Max Planck Institute for
Chemical Physics of Solids, 01187 Dresden, Germany}
\affiliation{Würzburg-Dresden Cluster of Excellence ct.qmat, Technische Universität Dresden, 01062 Dresden, Germany}
\author{A. Lubk}
\email{a.lubk@ifw-dresden.de}
\affiliation{Leibniz Institute for Solid State and Materials Research Dresden, Helmholtzstraße 20, 01069 Dresden, Germany}
\affiliation{Institute of Solid State and Materials Physics, TU Dresden, Haeckelstraße 3, 01069 Dresden, Germany}
%%%%%%%%%%%%%%%%%%%%%%%%%%%%%%%%%%%%%%%%%%%%%%%%%%%%%%%%%%%%%%%%%%%%%

\begin{abstract}

Disordered 2D lattices, including hexatic and various glassy states, are observed in a wide range of 2D systems including colloidal nanoparticle assemblies and fluxon lattices. Their disordered nature determines the stability and mobility of these systems, as well as their response to the external stimuli. Here we report on the controlled creation and characterization of a disordered 2D lattice of non-topological magnetic bubbles in the non-centrosymmetric ferrimagnetic alloy Mn$_{1.4}$PtSn. By analyzing the type and frequency of fundamental lattice defects, such as dislocations,  the orientational correlation, as well as the induced motion of the lattice in an external field, a non-ergodic glassy state, stabilized by directional application of an external field, is revealed. 

\end{abstract}

\maketitle
\section{Introduction}

The collective behavior of elastic manifolds in a random or rough background, along with the formation of glassy states, remains a central topic within the broader study of glasses, amorphous materials, and other disordered configurations of condensed matter. An early motivation for this research was the behavior of dislocations under pinning by mixed sites in alloy crystals \cite{labusch1970statistical}. These systems, often trapped in non-ergodic configurations, exhibit unique responses to external stresses, making their behavior particularly intriguing. The theories of dislocation-mediated melting in the context of the Berezinski-Kosterlitz-Thousless transition \cite{jose1977renormalization} provide a universal picture regarding the role of defect formation and plasticity due to quenched disorder, specifically for two-dimensional (2D) systems. Well studied examples in this context are vortex lattices in type-II superconductors \cite{Fisher1991,Blatter1994}, overlayers of colloidal particles \cite{Tierno2012,deutschlander2013two}, smectic \cite{bellini2001universality} and blue phases of liquid crystals \cite{henrich2010ordering}, or some phases of superfluidic helium \cite{osheroff1972evidence,volovik2008larkin,pollanen2012new}.

2D magnetic bubble lattices, and skyrmion lattices in gyroptropic non-centrosymmetric ferromagnets, provide another example of such states, where the assembly  of bubbles or skyrmions can be considered to form certain mesophases, intermediate between fully crystalline and liquid disordered states \cite{Reichhardt2022}. In 2D systems, the Mermin-Wagner theorem stipulates that fluctuations should preclude the formation of long-range ordered lattice states. However, in these magnetic systems, there are two effects promoting long-range order:  First, the classical dipole-dipole interactions, owing  to their long-range, and second, the magneto-crystalline anisotropy along with pinning in the regular crystalline lattice of the magnetic material. On the other hand, in a magnetic alloy, the atomistic disorder of the realstruktur creates a rough potential for the collective ordering of a magnetic states via random anisotropies and random exchange forces \cite{Kindervater2020}, which can counteract these tendencies towards long-range order. Hence, it is expected that magnetic bubble or skyrmion lattices in such a magnetic material behave quite differently compared to, e.g., a vortex lattice in a type-II superconductor, and allows to study different effects. 

Notably, quenched disorder and pinning determine the condensation of magnetic bubbles and skyrmions into a rich variety of (partly) ordered or disordered 2D lattices \cite{Reichhardt2016, Reichhardt2022}. While inter-skyrmion or inter-bubble interaction typically prefers condensation into a periodic triangular lattice \cite{Muehlbauer2009, Yu2010} (other regular lattices have been also observed \cite{Takagi2022}), quenching and random pinning can potentially lead to formation of glassy states (e.g., Bragg glass, vortex glass) \cite{Giamarchi1995}, as observed in lattices of other building blocks \cite{Fisher1998}, such as vortices in type-II superconductors \cite{AragonSanchez2019} or colloidal particles \cite{Pertsinidis2008}. More recently, it has been revealed that quenched disorder and pinning also determine the response (both static and dynamic) of magnetic skyrmions and bubbles under application of external stimuli (magnetic fields, electrical currents, temperature (gradients)) \cite{Litzius2020}. While there is a considerable number of observations of disordered 2D lattices of magnetic skyrmions and bubbles in thin films (e.g., \cite{Hsu2018, Wang2019}), research mainly focused on interplay with magnetic frustration \cite{Karube2018}, geometric boundaries such as domain walls \cite{Nakajima2017}, characterization of particular lattice defects such as grain boundaries \cite{Matsumoto2016}, as well characterization of the hexatic phase \cite{huang2020melting}, while glassy states in skyrmion or bubble lattices remain less well studied.

Due to its rich phenomenology of magnetic structures and the intrinsic proliferation of crystallographic defects, the non-stochiometric inverse half-Heusler compound Mn$_{1.4}$PtSn offers a suitable platform to study disordered lattices. Mn$_{1.4}$PtSn crystallizes in the space group $\mathrm{I\bar{4}2d}$, and has been the focus of extensive research efforts for its multitude of magnetic textures including lattices of magnetic solitons \cite{Nayak2017}. The non-centrosymmetric structure of Mn$_{1.4}$PtSn gives rise to anisotropic Dzyaloshinskii-Moriya interaction (DMI) as well as a sizable uniaxial anisotropy. The competing magnetic interactions result in a large variety of magnetic textures, including non-topological bubbles, elliptical skyrmions and antiskyrmions \cite{Nayak2017, Peng2020, Jena2020}, that may be transformed into each other by applying (a sequence of) external magnetic fields of particular direction and strength. The large number of atomic vacancies and other lattice defects in Mn$_{1.4}$PtSn provides a dense background of pinning sites, with their individual extension being small with respect to that of the magnetic textures in that material. Therefore, pinning is determined by the variation of the defect background rather than by individual pinning sites, which may be described by a slowly varying pinning potential.

In the following, we reveal a mechanism that exploits the rich landscape of metastable textures in combination with the random background potential to deliberately create a disordered lattice state in Mn$_{1.4}$PtSn. Subsequently, we characterize this bubble lattice state in terms of lattice defects, spatial correlations, and response to an external magnetic field. 

\section{Creation of disordered non-topological bubble lattice}
Non-stochiometric Mn$_{1.4}$PtSn single crystals have been synthesized by a self-flux method as described in Ref.~\cite{Vir2013}. Subsequently, a thin, electron-transparent, lamella of approximately 100 nm thickness (lamella normal parallel to the $c$-axis) has been prepared by focused ion beam (FIB) milling. The magnetic textures in the thin lamella were subsequently investigated with a JEOL F200 Transmission Electron Microscope that was operated in Lorentz mode (Lorentz Transmission Electron Microscopy -- LTEM). In this mode the objective lens is largely switched off to prevent field magnetization along the optical axis of the transmission electron microscope and the specimen plane is slightly defocused to observe magnetic phase contrast. 

\begin{figure}[h]
   \centering
    \includegraphics[width=\linewidth]{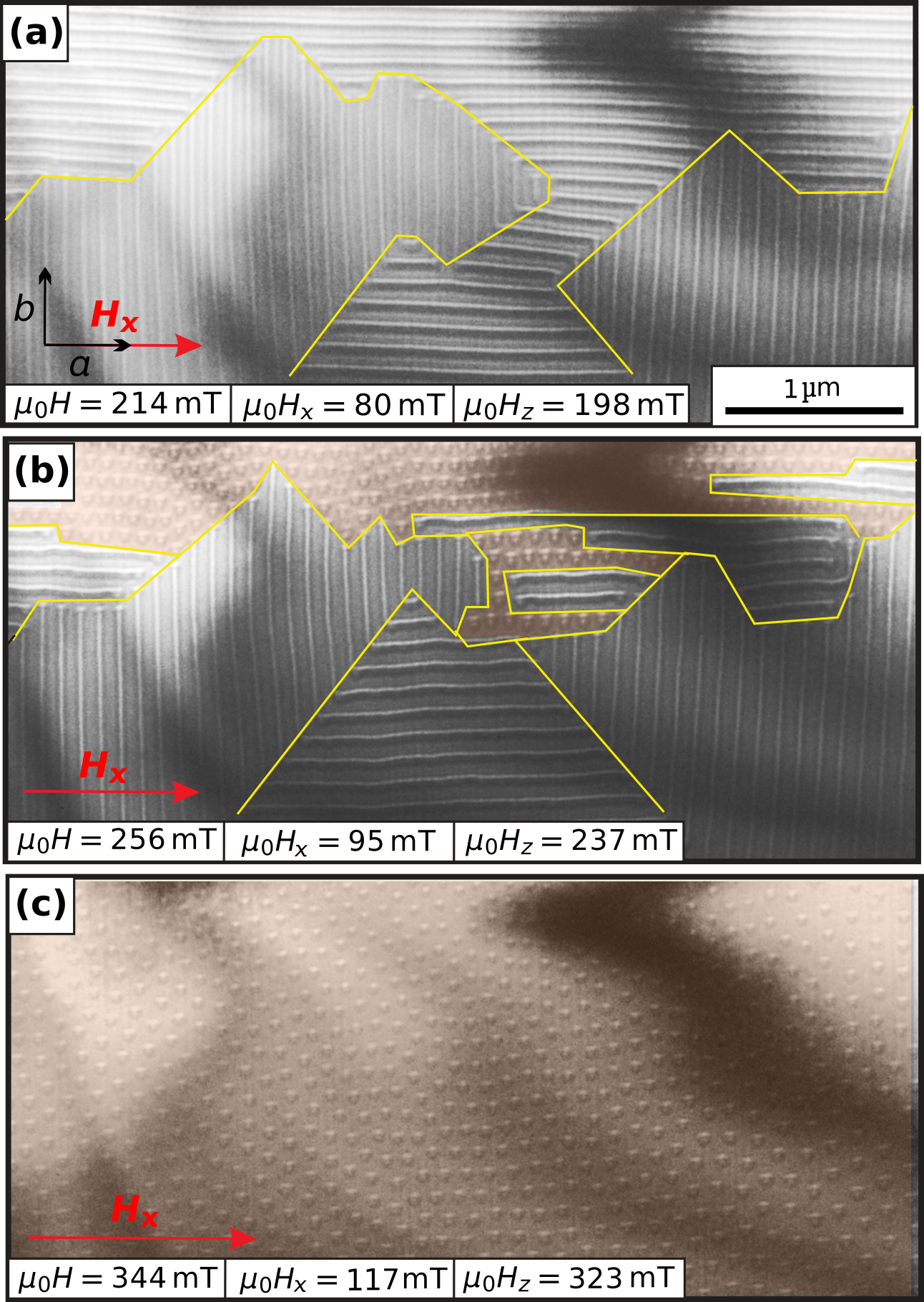}
   \caption{The disordered bubble lattice state is stabilized by the application of increasing magnetic fields under an oblique angle of $\theta=20\degree$. The field strengths are indicated as the full field strength $\mu_0H$, and the respective components $\mu_0H_x$ and $\mu_0H_z$. (a) A mixed domain state of single-Q magnetic textures propagating along the $a$- and $b$-axis, respectively, is observed with a total applied field strength of $\mu_0H=214\,\mathrm{mT}$. The borders of the domains are indicated by yellow lines. (b) At $\mu_0H=256\,\mathrm{mT}$, the domain with its propagating vector aligned perpendicular to the in-plane field $\vect{H_x}$ partially transitions into a disordered bubble lattice. (c) Full sample coverage of the disordered bubble lattice is achieved at $\mu_0H=344\,\mathrm{mT}$.}
   \label{fig:bubble_lattice_creation}\end{figure}

2D lattices of magnetic bubbles have been created in the lamella by applying increasing magnetic fields (field direction fixed along the optical axis), while deliberately tilting the sample $\theta=20\degree$ away from the $c$-axis around the $b$-axis. This allows adjustment of in- and out-of-plane magnetic fields with respect to the lamella. Upon increasing the external field, the initial helical state partially transforms into an intermediate homochiral state (Fig. \ref{fig:bubble_lattice_creation}(a)). Subsequently, the initial helical state is partially transformed into a bubble state (Fig. \ref{fig:bubble_lattice_creation}(b)). A complete magnetic bubble lattice appears above $344 \, \mathrm{mT}$ as depicted in Fig. \ref{fig:bubble_lattice_creation}(c). In this transformation process, the disorder is introduced by the formation of irregular boundaries as well as spatially separated nucleation sites. The precise characterization of the intermediate states, as well as the boundaries between different magnetic textures, is the subject of a separate study.

\begin{figure}[h]
    \centering
 \includegraphics{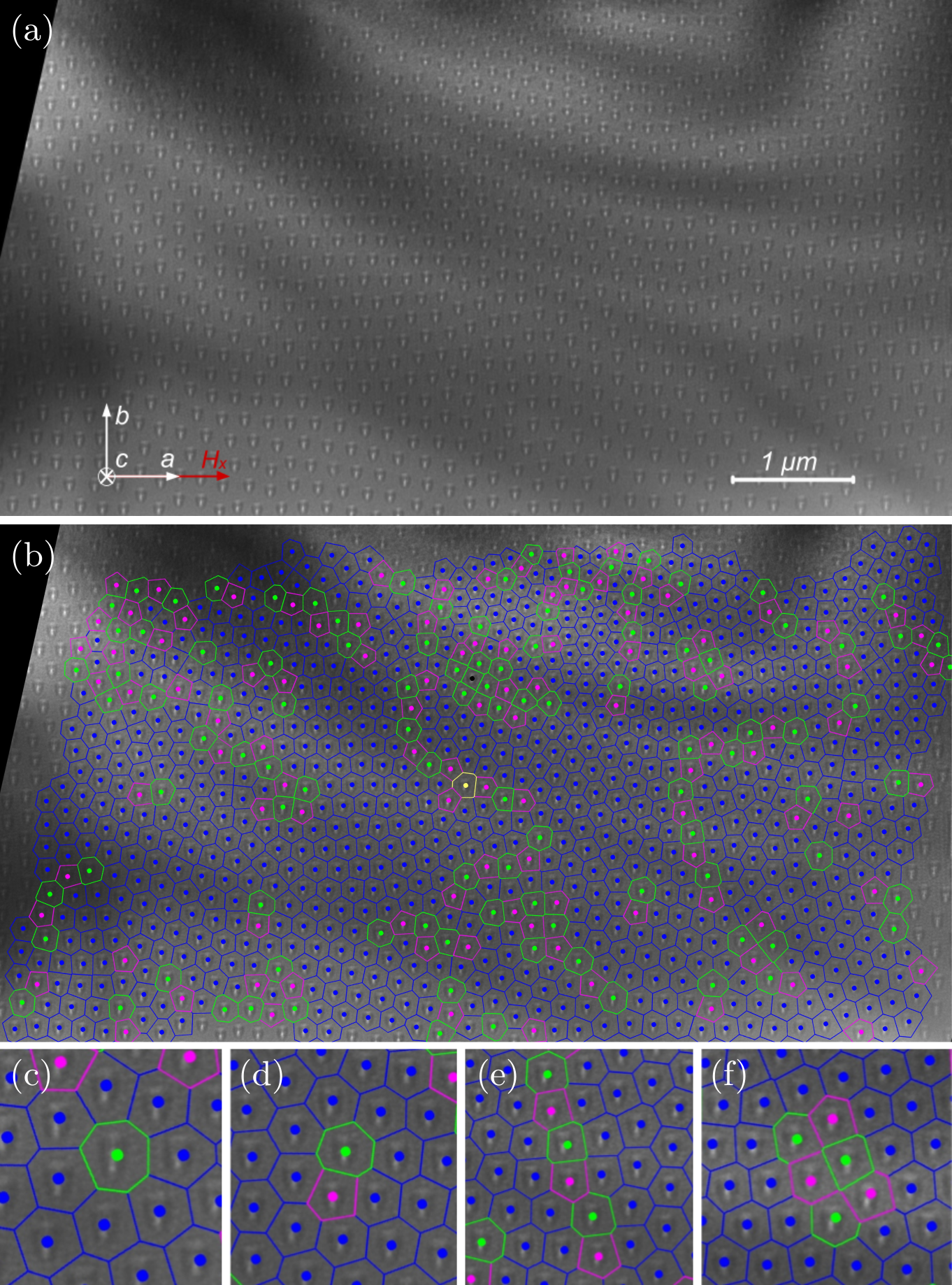}

\caption{(a) LTEM image of a 2D lattice of topologically trivial magnetic bubbles in Mn$_{1.4}$PtSn observed at room temperature in an applied field of $344 \, \mathrm{mT}$ under an angle of $\theta=0.9\degree$ with respect to the $c$-axis. The crystal orientation and in-plane projection of the magnetic field due to the slight tilt of the sample around crystallographic $a$-axes are indicated. (b) Non-topological bubble lattice in Mn$_{1.4}$PtSn. Solid dots denote the positions of the bubbles extracted from (a). The lattice sites are colored according to the number of nearest neighbors using the following scheme:  4--black, 5--pink, 6--blue, 7--green,  8--yellow. The dominant deviation from the six-fold coordinated trigonal packing are 5-fold and 7-fold coordinated defects. The bottom row shows zoom-outs of lattice defects within the generally hexagonal bubble lattice: (c) standalone 7 neighbor bubble (disclination),  (d) two bound disclinations with 5 (pink) and 7 (green) neighbors forming a dislocation, (e) chain of dislocations forming a grain boundary, (f) twisted bond (2 bound dislocations) bound to a dislocation.
}
\label{fig:MnPtSn_TEM}
\end{figure}

Subsequently, to improve the LTEM contrast, the tilting angel was reduced to $\theta=0.9\degree$ while maintaining the applied field strength. The bubble lattice created by this protocol is relatively sparse, as evident in Fig. \ref{fig:MnPtSn_TEM}(a), with an average skyrmion distance of $~0.18\,\mu m$, significantly exceeding their individual sizes. As a result, their mutual interaction is primarily governed by long-range dipolar forces and rather weak, which is generally promoting the formation of defects. Moreover, disorder does not impact the internal structure of the skyrmions or bubbles at lattice defects as, e.g., observed in densely packed lattices or defects \cite{Matsumoto2016}.

\section{Characterization of the Disordered Bubble Lattice}

In order to characterize the lattice, we determined the bubble positions in a first step by employing a pattern recognizing neural network. Partially annotated data from another experiment, featuring similar bubbles, was utilized for network training. Given the limited annotated dataset, transfer learning was employed \cite{yang2020transfer} utilizing the pre-trained Efficient Det D2 768x768 model from the TensorFlow library. To detect new custom objects, we froze the section of the model responsible for identifying "features" in the image and fine-tuned the layers responsible for classifying those features into the desired objects. We furthermore notate that the annotated training dataset was sparse, i.e., not all existing objects in the images were annotated. Thus, to facilitate better learning, all regions potentially containing unannotated objects were filled with zero values. This was implemented to prevent the model from taking into account potentially misannotated areas. The image for bubble detection was significantly larger than the training images. Therefore, a sliding window approach was employed. This involved splitting the input image into smaller segments, and the detection results from the model were aggregated into a unified outcome.

The result of this step is displayed in Fig. \ref{fig:MnPtSn_TEM}(b), where 6-fold coordinated bubbles (standard case in triangular lattice) are colored blue, 5-fold pink, and 7-fold green. Upon inspection of the coordination number, we observe that isolated 5- or 7-fold coordinated bubbles, which correspond to disclinations (see Fig. \ref{fig:MnPtSn_TEM}c), are rarely present. Bound 5-7 defect pairs, i.e., dislocations (Fig. \ref{fig:MnPtSn_TEM}d), on the other hand, are occasionally observed within the lamella. Their proliferation (and the absence of disclinations) is characteristic for the hexatic phase \cite{huang2020melting}. When arranged as chains, dislocations form grain boundaries (Fig. \ref{fig:MnPtSn_TEM}e). The dislocation density along the one-dimensional trace of the boundary determines character and angular mismatch of the adjacent well-ordered domains or grains. Straight and bent domain boundaries are observed throughout the lamella. Finally, we also observe twisted bonds composed of two adjacent edge dislocations (or more) with opposite Burgers vectors (Fig. \ref{fig:MnPtSn_TEM}f). These lattice configurations correspond to local elastic deformations that typically occur in connection with grain boundaries.

\begin{figure}
\centering
\includegraphics{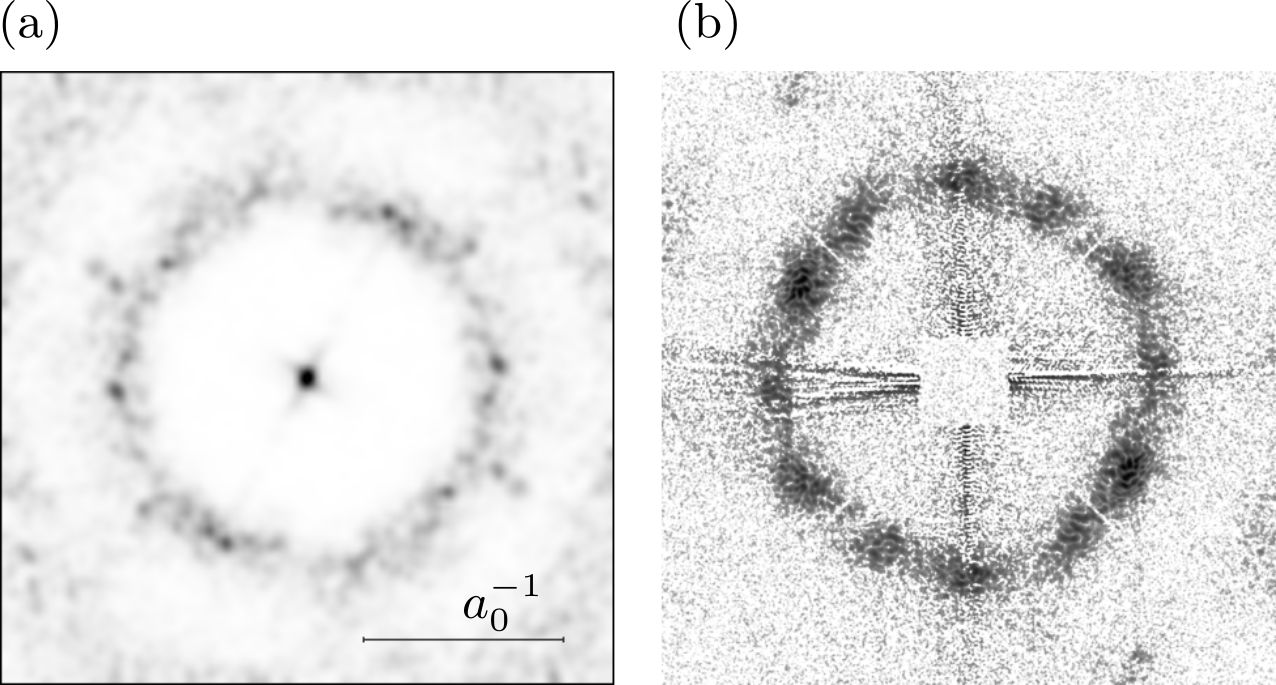}
\caption{(a) Structure factor of disordered bubble lattice computed from Eq. (\ref{eq:S}). (b) Transmission REXS diffraction pattern of a different lamella of similar geometry and in a similar external magnetic field. Note that the white lines and the white square are due to the beam stop support.}
\label{fig:MnPtSn_structure_factor}
\end{figure}

Another perspective on the 2D lattice state is obtained by analyzing the structure factor of the whole field of view (Fig. \ref{fig:MnPtSn_structure_factor}a). In the following the structure factor is computed from the bubble positions $\mathbf{R}_j$ by evaluating
\begin{equation}
    S\left(\mathbf{k}\right)=\frac{1}{N}\left|\sum_{j=1}^Ne^{-i\mathbf{k}\mathbf{R}_j}\right|^2\,,\label{eq:S}
\end{equation}
where $N$ denotes the total number of bubbles and $\mathbf{k}$ is the reciprocal lattice vector. In this definition the structure factor corresponds to the recorded intensity in a diffraction experiment on point-like scatterers \cite{warren1990x}.
The structure factor reveals a diffuse ring of first order, and a faintly visible one of second order. For comparison, we also show the diffracted intensity from resonant elastic x-ray scattering experiments (REXS) carried out on lamellae under external field conditions. This REXS experiment has been carried out on the I10 beamline at DIAMOND Light Source (UK), using the portable octupole magnet system (POMS) \cite{VANDERLAAN201495}. The photon energy was tuned to the Mn $L_3$-edge ($642.2\,\mathrm{eV})$. Due to the wavelength needed for resonant scattering as compared to the magnetic lattice spacing (momentum transfer in the first Brillouin zone), the experiment was carried out in transmission. The observed absence of systematic reflections in the structure factor corresponds to a large orientational disorder within the field of view (the faintly visible presence of local maxima may be related to finite field of view effects and hence insufficient statistics).

From these observations we may already conclude that the lattice under consideration is not a completely disordered liquid. On the other hand, the lattice is not well-ordered either (i.e., no Bragg glass). We may, however, also exclude the hexatic phase, which would exhibit a smaller orientational disorder (e.g., visible through distinguishable systematic reflections). Furthermore, the distribution of defects does not correspond to a multigrain state composed of well-defined (i.e., separated by grain boundaries) grains of different orientation. These findings point towards the presence of a glassy state.

In order to further elucidate the nature of the observed lattice state, we elaborate on the orientational correlation function in the following. 

For a 2D hexagonal lattice, the local orientational order parameter can be defined through the bond orientations between nearest-neighbouring sites \cite{jose1977renormalization}:
\begin{equation}
\Psi_{6}(\textbf{r}_i) = \frac{1}{N_{nn}} \sum_{j}^{N_{nn}} e^{i6\Theta_{ij}}\,,
\end{equation}
where ${N_{nn}}$ represents the number of nearest-neighbor bubbles around the reference particle located at position $\textbf{r}_i$, which can be determined by Delaunay triangulation. $\Theta_{ij}$ is the angle between the
$ij$ bond and an arbitrary but fixed axis. The
orientational correlation function $G_6(r)$ is then
\begin{equation}
G_6(r) = \frac{1}{N_r} \sum_{(i,j)}^{N_r} \Psi_{6} (\textbf{r}_i) \Psi_{6}^* (\textbf{r}_j)\,,
\end{equation}
where ${N_r}$ is the number of bubbles at distance $r=|\textbf{r}_i-\textbf{r}_j|$.

\begin{figure}
    \centering
    \includegraphics{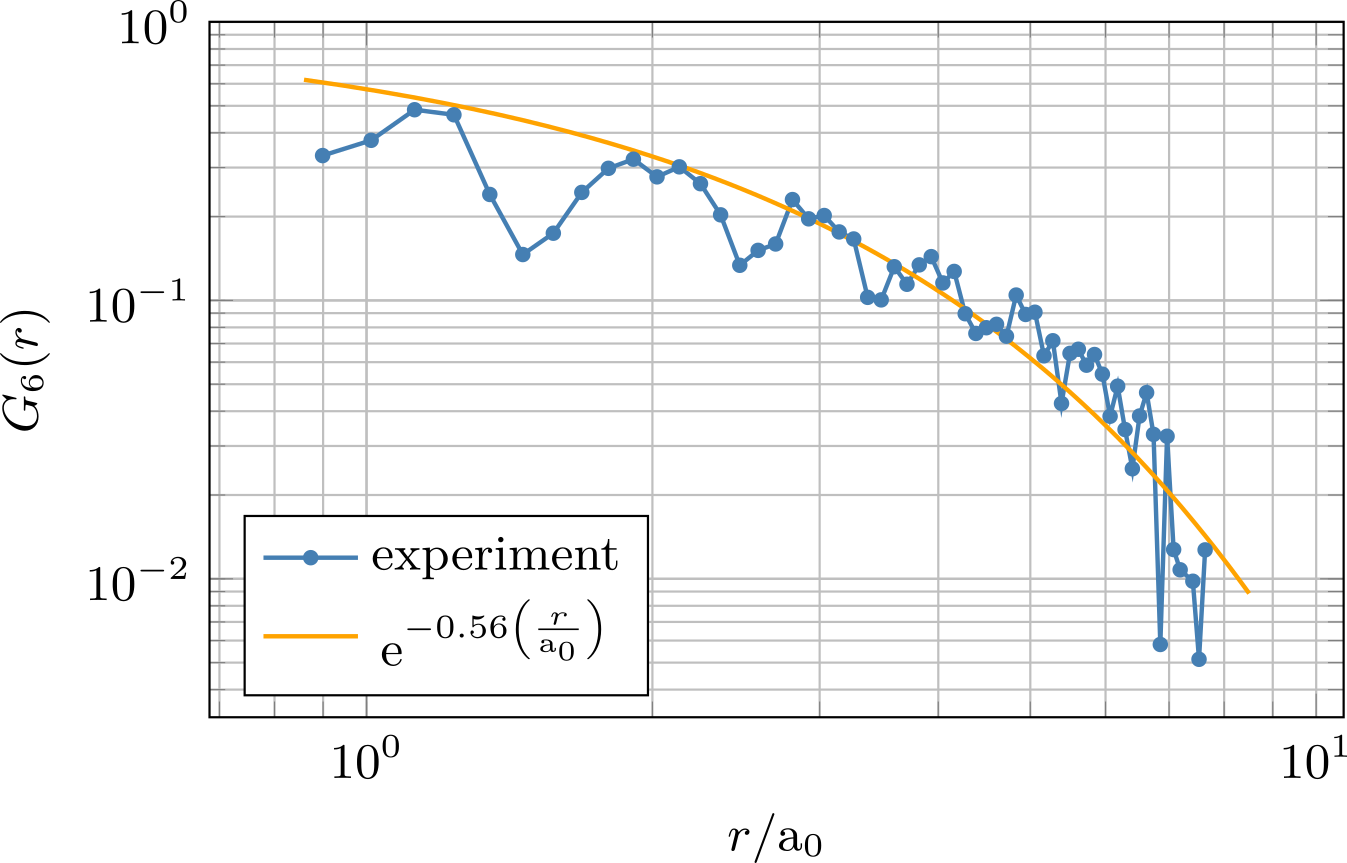}
    \caption{Orientational correlation function $G_6(r)$ of the magnetic bubble lattice as a function of the normalized distance $r/a_0$ with $a_0$ being the mean bubble distance (blue curve). The orange curve represents an exponential fit to the data neglecting the local minima results in $G_6(r)=\text{e}^{-0.56\left(\frac{r}{a_0}\right)}$.}
\end{figure}

The correlation function exhibits an exponential instead of an algebraic decay. This short range orientational correlation corroborates that the observed phase is not hexatic (exhibiting algebraic long-range orientational correlation). Moreover, the absence of long-range orientational correlation implies the absence of long-range translational correlation. Indeed, vortex glass phases previously observed in superconducting vortex lattices \cite{AragonSanchez2019}, show a similar short range orientational order. The origin of the latter is the prevalence of dislocations, typically forming grain boundaries and twisted bonds. To gain further insight we finally study how a change in the external magnetic field is deforming the observed lattice state.

\section{Glass-like motion}

\begin{figure}[h]
    \includegraphics{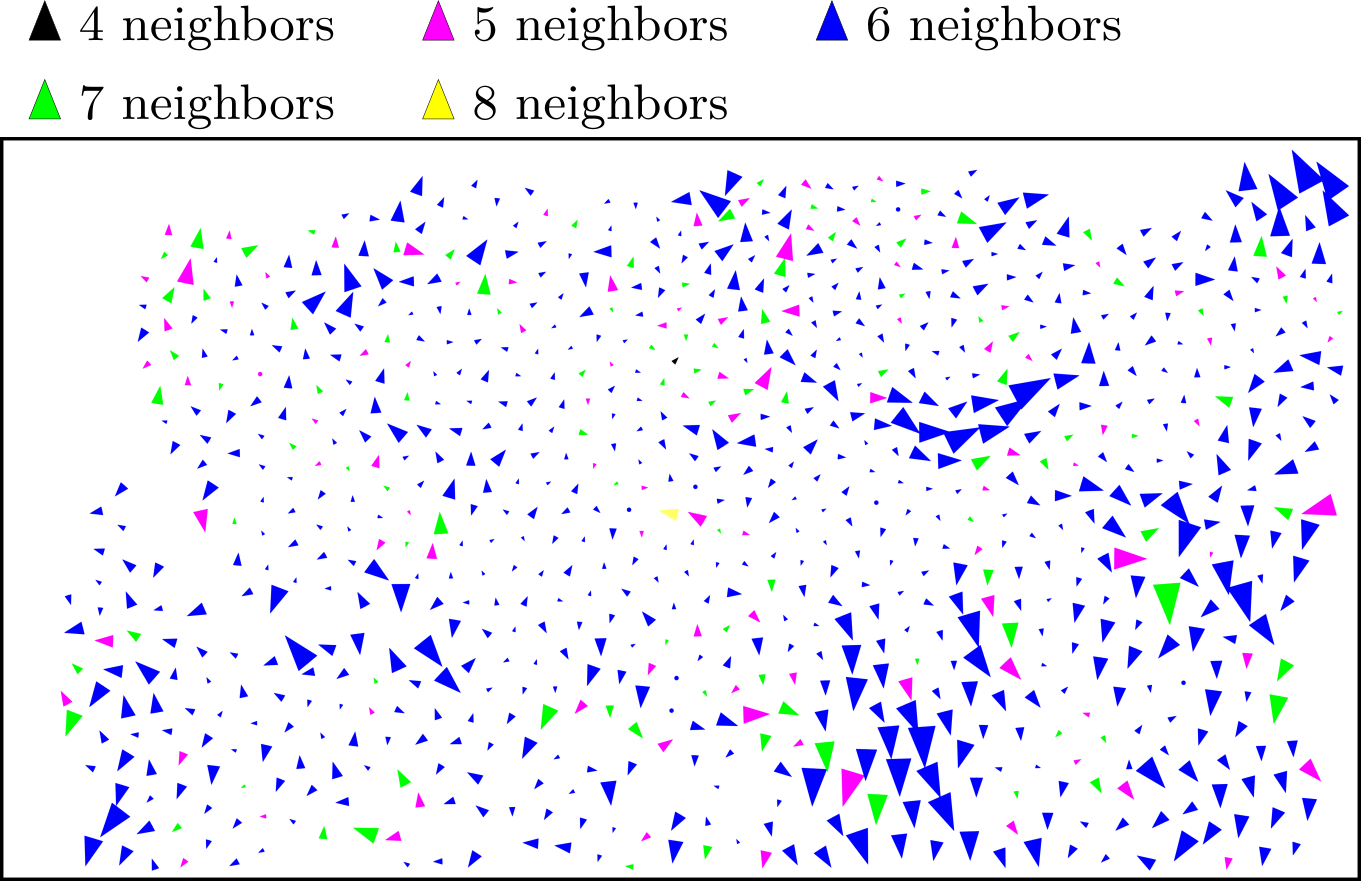}
    \caption[width=\textwidth]{Displacement of bubbles of the lattice shown in Fig.\,2 upon slight variation of external magnetic field angle (by tilting the lamella). The size of the arrows is proportional to the displacement. Bubbles with six neighbors are depicted with blue arrows. Bubbles with five and seven neighbors are shown in purple and green, respectively. Black and yellow arrows correspond to four and eight neighbours respectively.}
    \label{fig:MnPtSn_motion}
\end{figure}

As discussed above, the non-topological bubble lattice in Mn$_{1.4}$PtSn can be prepared by increasing the external magnetic field in a carefully chosen orientation. The resulting sequence of magnetic transformations of magnetic textures leads to a non-ergodic state and rather large modulations of intermediate states. In the following, we study the impact of a small change of the external field. The small field change is facilitated by tilting the lamella by $0.8\degree$. By analyzing the bubble positions, we identified their displacement resulting from this slight field change (Fig. \ref{fig:MnPtSn_motion}). The observed motion is strongly confined to particular regions of the lattice and could not be reverted by returning to the original tilt. Such a plastic motion is indicative of a glassy state.  In strongly disordered 2D lattices, such as glass states, particles typically depin plastically, leading to locally large distortions and shifts, while others remain pinned \cite{Reichhardt2022}. The depinning typically occurs in the vicinity of defects such as grain boundary edges. Indeed, we observe a weak correlation of plastic motion and defects in regions of small defect density that are more stable than regions of large density. Note, however, that the small sample size does not allow a statistically significant corroboration of that observation.

\section{Summary}

In summary, we have demonstrated a pathway for creating strongly disordered 2D lattices of non-topological magnetic bubbles in Mn$_{1.4}$PtSn. Herein, the disorder is imprinted by subsequent nucleation of different magnetic textures in an external magnetic field that are spatially separated by irregular boundaries. The shape and formation of these boundaries is presumably influenced by the random background pinning potential originating from the large number of atomic lattice defects in this non-stochiometric compound. The bubble lattice shows characteristics of a glass state in the locally observed lattice defects, the structure factor, and the exponential decay of the orientational correlation. 

These results shed light on the general phenomenology of disordered magnetic skyrmion and bubble lattices, as pinning of the condensed nanoscale solitons by alloying disorder will, to some extend, always be present in real alloyed chiral helimagnetic systems. The induced plastic behavior of the lattice may be detrimental to applications aiming for the fast translation of bubble or skyrmion-based information units. On the other hand, stable storage of information may require that many different metastable configurations can be created and maintained over long times. Therefore, next steps towards such applications should concentrate on exploring dynamic response at various timescales, and developing means to manipulate the observed glassy state in Mn$_{1.4}$PtSn.

\section{Acknowledgements}
S.P.\ acknowledges support through the Philipp Schwartz Initiative of the Humboldt Foundation.
M.W.\ acknowledges support from the International Max Planck Research School for Chemistry and Physics of Quantum Materials (IMPRS-CPQM).
M.C.R.\ was supported through the Emmy Noether programme of the German Science Foundation, DFG, (project no.\ 501391385). A.L. acknowledges support by the DFG through CRC 1143. C.F. acknowledges support by the Cluster of Excellence ct.qmat (EXC 2147, project ID 390858490).
M.W., D.P and B.R.\ are grateful for funding from the DFG through SPP 2137, project no.\ 403503416. J.S. received funding from the HORIZON EUROPE framework program for research and innovation under grant agreement n. 101094299.
REXS was conducted on the Portable Octupole Magnet System on beamline I10 at the Diamond Light Source, UK, under proposal MM28882. Financial support from the UK Skyrmion Project (Engineering and Physical Sciences Research Council, EP/N032128/1) is gratefully acknowledged. J.B. acknowledges a Diamond-EPSRC studentship (2606404, EP/R513295/1, EP/T517811/1).

\bibliographystyle{apsrev4-2}
\bibliography{bib}

\end{document}